\documentclass[twocolumn,tighten]{aastex62}

\usepackage{placeins}
\usepackage{natbib}
\usepackage{amsmath}
\usepackage{color}
\usepackage{enumerate}
\usepackage{graphicx}
\usepackage{soul}

\shorttitle{Angular momenta of protostellar cores}
\shortauthors{Kuznetsova et al.}

\begin{document}


\title{The Origins of Protostellar Core Angular Momenta}

\correspondingauthor{Aleksandra Kuznetsova}
\email{kuza@umich.edu}

\author{Aleksandra Kuznetsova}
\affil{Department of Astronomy, University of Michigan, 1085 S. University Ave., Ann Arbor, MI 48109}
\author{ Lee Hartmann}
\affil{Department of Astronomy, University of Michigan, 1085 S. University Ave., Ann Arbor, MI 48109}
\author{Fabian Heitsch}
\affil{Department of Physics and Astronomy, University of North Carolina - Chapel Hill}


\begin{abstract}
We present the results of a suite of numerical simulations
designed to explore the origin of the angular momenta of protostellar cores. Using the 
hydrodynamic grid code \emph{Athena} with a sink implementation, we follow the formation of protostellar cores and protostars (sinks) from the subvirial collapse of molecular clouds on larger scales to investigate the range and relative distribution of core properties.  We find that the core angular momenta are relatively unaffected by large-scale rotation of the parent cloud; instead, we infer that angular momenta are mainly imparted by torques between neighboring mass concentrations and exhibit a log-normal distribution. 
Our current simulation results are limited to size scales $\sim  0.05$~pc ($\sim 10^4 \rm AU$), but serve as first steps toward the ultimate goal of providing initial conditions for higher-resolution studies of core collapse to form protoplanetary disks
\end{abstract}


\citestyle{aa}

\section{Introduction}

Disk formation is a natural consequence of gravitational collapse of protostellar cores even without complete angular momentum conservation, given the large difference in sizes between cores and stars.  The disk mass surface
density distribution, of obvious importance for
understanding planet formation, results
from the angular momentum distribution of the infalling protostellar envelope, modified by transport processes within the disk \citep{cassen81}.  The expected low turbulence in large regions of
protoplanetary disks \citep[see review by][]{turner14}, supported by observational limits \citep{flaherty17,flaherty18} suggest that turbulent angular momentum transport is generally quite slow.  Transport by
disk winds may dominate \citep{bai16}, but winds may well be trapped by the infalling envelope, rendering them ineffective in
redistributing mass until the infall phase ends.  
Thus, the early structure of protostellar and protoplanetary disks may be dominated by the angular momentum distribution of the parent infalling envelope. Given the increasing evidence for early planet formation \citep[e.g.,][]{alma15}, developing a better 
understanding of envelope angular momenta is essential.

The analytic rotating collapse model of
\citet{TSC_84} (hereafter TSC) has been
used often to predict disk structures and other properties assuming
various levels of turbulent viscosity
\citep[for example,][]{zhu10,bae14,bae15,hartmann18}. However, the assumption of initial solid-body rotation in the TSC
model is not necessarily realistic; differing distributions of angular momenta would result in differing initial disk mass distributions.

\par Due to the short dynamical timescales and spatial orders of magnitude inherent in gravitational collapse, a top-down approach that starts from cluster scales is necessary to develop better parameters for model input that accurately reflect that of a population of protostellar disks formed in a realistic environment.  By far the most
extensive investigation of disk formation
with this approach is
that of \cite{Bate_2018}, who analyzed the properties of circumstellar disks formed in a radiation hydrodynamic simulation \citep{bate12}. The disks exhibited a wide range of properties, with typical radii $\sim 100$~AU and surface densities $\Sigma \propto R^{-1}$, in reasonable agreement with
observational constraints.
However, many disks were not well-resolved, and the amount of disk
evolution due to artificial viscosity 
was not clear.

In this paper we address a simpler, more basic, and
more self-contained problem: what sets the angular momenta of protostellar cores?  To this end, 
we present results from a series of numerical experiments based on the picture
of cluster formation via sub-virial (cold) collapse, as in \cite{Kuznetsova_2015}.  These simulations provide a distribution of initial conditions for disk formation by core collapse and allow an investigation of the importance of global cloud rotation on the angular momenta of
individual cores.

Using a sink implementation that keeps track of properties in the near environments of sinks (e.g., the surrounding cloud core), we find that the resulting  angular momenta do not behave according to the expectations of the TSC model, with no detected smooth growth of angular momentum over time for the protostars as a whole.  This suggests that initial disk surface densities might be flatter, i.e. a weaker function of radius, than typical disk models, with implications for planet formation. The angular momentum of cores is insensitive to the global cloud rotation, indicating that the angular momentum inputs to the
cores are
the result of local gravitational torques.  As in \cite{Bate_2018}, we find considerable time variability in the accretion of the angular momenta and mass, which is the product of a lumpy episodic non-isotropic accretion of material onto the protostellar cores.  These first results set the stage
for further studies exploring the role
of magnetic fields and higher-resolution simulations to provide more detailed collapse models as an essential input to
investigations of protoplanetary disk structure.

\section{Method}

\subsection{Basic Assumptions and Sink Implementation}
\par 
The methods used here were described in  \citep{Kuznetsova_2018b}, so we provide a short summary here.
We use  a modified version of the Eulerian grid code \emph{Athena} \citep{Stone_2008} to simulate the collapse of a molecular cloud by self-gravity.
We solve the system of equations
\begin{eqnarray}
\label{eq:mass}
\frac{\partial\rho}{\partial t} + \nabla \cdot (\rho \mathbf{v}) = 0\\
\label{eq:mom}
\frac{\partial \rho \mathbf{v}}{\partial t} + \nabla \cdot (\rho\mathbf{vv} + P) = -\rho \nabla \Phi\\
\label{eq:poisson}
\nabla \Phi = 4\pi G \rho
\end{eqnarray}
with an RK3 integrator \citep{GottliebShu1998}, which advances the fluid equations (eqs. \ref{eq:mass} and \ref{eq:mom}) at third order in time. 
We further adopt an isothermal equation of state such that $P=c_s^2 \rho$ for simplicity, which is a reasonable approximation for low-mass star-forming regions on the scales we study. The Poisson equation (eq.~\ref{eq:poisson}) is solved every RK3 substep, using the FFT solver that comes with version 4.2 of Athena. 

\par Using the same methods as our preceding companion paper on the accretion of the cores \citet{Kuznetsova_2018b},  we adopt a sink-region geometry similar to those described in \citet{Bleuler_2014, Gong_2012}, where at every timestep an accretion reservoir is drawn centered on the sink, which will hereinafter be referred to as the \emph{sink-patch}. The radius of the sink-patch is set by the parameter $r_{acc}$, which describes the number of cells to be included in the patch radius, in addition to the central cell which houses the sink particle. Sink accretion of radially infalling material occurs instantaneously across the patch and leftover angular momentum is deposited in the patch cells; the diameter of the sink-patch, $2r_{acc} + 1$ cells , is the relevant resolution element to consider for our study.

\subsection{Initial Setups}
All of the simulations are initialized in a 20~pc box with a spherical top-hat density profile, where the ambient density is
$\rho_0 = 1.5 \times 10^{-23}$ g $\mathrm{cm^{-3}}$ and the 4 pc radius spherical cloud has a density of $\rho_c = 1.5 \times 10^{-21}$ g $\mathrm{ cm^{-3}}$, giving an initial free-fall time of $t_\mathrm{ff} = 1.7$ Myr for the cloud. 
The initial conditions are seeded with a decaying Mach 8 supersonic turbulent velocity spectrum $P(k) \propto k^{-4} dk$, which introduces some base level of cloud angular momentum. On average, across all the turbulent random seeds, the total initial angular velocity from turbulence at the cloud scale is $\Omega_k = 0.1 \pm 0.03 \ \rm km s^{-1} pc^{-1}$, where the specific angular momentum at the cloud scale about the central axis from the cloud scale eddies is $j = 1.97 \times 10^{23} \pm 0.6 \ \rm cm^2 s^{-1}$. For greater physical insight, in the following we specify $\Omega^{-1}$ values in terms of Myr. We explore a parameter space of additional angular momentum input with several values of constant angular velocity added to the entire cloud which we refer to by the initial rotation period for the entire cloud: $\Omega_c^{-1} = 6, 3, 1.5 \, \rm Myr $ (compare to the initial free-fall time of cloud at $1.7 \  \rm Myr$), where the turbulence only runs have average cloud scale rotation periods of $\Omega_k^{-1} \sim 10 \ \rm Myr$. We supplement the runs in \citet{Kuznetsova_2018b} with additional high resolution runs, such that there is data at three resolutions; $N_{\ cell} = 256^3, 512^3, 1024^3$ and where the cell size is then $\Delta x = 0.08, 0.04, 0.02 \, \rm pc$, respectively. In this paper, we focus on data from the intermediate and highest resolutions to compare resolution effects and discuss the measurement of angular momentum at different scales. The fiducial run used in this work, HR\_s2, has $1024^3$ grid cells with $\Delta x = 0.02 \, \mathrm{pc}$,
a patch radius of $r_p = (r_{acc} + 1/2)\Delta x = 0.05 \, \mathrm{pc}$, and an isothermal temperature $T=14 \ \mathrm{K}$. The list of runs used in this work and some of their attributes can be found in Table \ref{tab:allruns}. 

\begin{table*}[ht!]
\caption{A tabulated list of runs and their initial parameters used for this study. All runs have $r_{acc} = 2$ and $T= 14 \ \rm K$. }
\begin{tabular*}{\textwidth}{c @{\extracolsep{\fill}} lccrrrr}
\hline
Run        & Seed & $\Omega_c^{-1}$ {[}Myr{]}& $\Omega_k^{-1}$ {[}Myr{]}& $\rm t_{end}$ {[}Myr{]} & $\rm N_{sink}$ & $\langle n_s \rangle \rm \ {[}pc^{-3}{]}$  \\ \hline
\multicolumn{5}{l}{$N_{cell} = 512^3$}                                         \\ \hline
IR\_s0     & 0    &  ...  &      15           & 2.35                & 98   &   0.018   \\ 
IR\_r1\_s0 & 0    & 6     &      15               & 2.35                & 66 & 0.012        \\ 
IR\_r2\_s0 & 0    & 3     &      15                   & 2.35                & 42    & 0.009     \\ 
IR\_r3\_s0 & 0    & 1.5   &      15                 & 2.35                & 19  &  0.002     \\ \hline
\multicolumn{5}{l}{$N_{cell} = 1024^3$}                                        \\ \hline
HR\_s2     & 2    & ...  &        10               & 2.1                 & 115  &   0.028   \\ 
HR\_s0     & 0    & ...  &         15           & 1.9                 & 110      &  0.024 \\ 
HR\_s1     & 1    & ...  &         8           & 2.4                 & 70        &  0.011 \\ 
HR\_r1\_s0 & 0    & 6    &        15            & 1.9                 & 122      &  0.028\\ 
HR\_r1\_s1 & 1    & 6    &          8          & 2.5                 & 88        &  0.014\\ 
HR\_r2\_s0 & 0    & 3    &          15           & 1.9                 & 134     &  0.032 \\ 
HR\_r2\_s1 & 1    & 3    &           8          & 2.5                 & 54       &  0.009\\ 
HR\_r3\_s0 & 0    & 1.5  &           15          & 1.9                 & 102     &  0.024 \\ 
HR\_r3\_s1 & 1    & 1.5  &            8         & 2.5                 & 10       &  0.001\\ \hline
\end{tabular*}
\label{tab:allruns}
\end{table*}

\subsection{Sink-Patch Data}
\par Sink-patch data is output as the three dimensional sink velocity, sink position, and the conserved variables in the patch cells (and an additional boundary cell). The angular momentum of the patch is summed over the entirety of the patch cells $L = \Sigma_{i} m_i \mathbf{v}_i \times \mathbf{r}_i$ where $m_i$ is the mass enclosed within the cell, $\mathbf{v}_i$ is the cell velocity relative to the sink and $\mathbf{r}_i$ is the radial distance of the cell center from the sink. The specific angular momentum is then the angular momentum divided by the total mass enclosed within the patch radius. As mass is removed from cells during sink accretion, angular momentum is removed from the patch and
is implicitly put into the sink. The sink angular momentum is tracked during the calculation and is always an insignificant fraction of the patch angular momentum, so we do not consider it further. 

\par To ensure that we discuss the angular momentum inheritance and evolution of systems where the angular momentum will go into the disk, we filter the dataset to remove sinks that are not likely to be single systems by virtue of three possible processes detailed below; sink merging, unresolved fragmentation, or rotational fragmentation.

\begin{figure*}[htb]
    \centering
    \includegraphics[width=\textwidth]{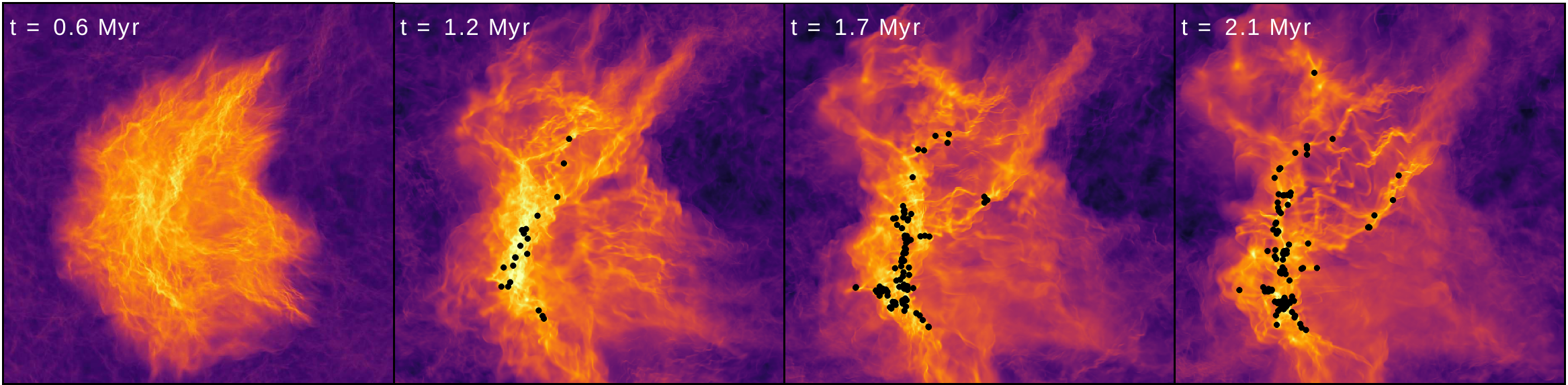}
    \caption{The evolution of the cloud in the high resolution ($N_{cell} = 1024^3$) fiducial run, shown here as the column density in the inner 12 pc of the simulation domain at four different snapshots in time $t = 0.6, 1.2, 1.7, 2.1$ Myr.}
    \label{fig:mainevol}
\end{figure*}

\par Sink merging occurs when one sink enters anothers patch; these sinks are then merged into one sink with a combined mass. We identify sinks by unique id numbers and remove all instances of sinks post merger in the data set. The sink implementation requires both a negative gradient in the potential and an increasing density profile across the entire patch, so fragmentation due to the presence of two perturbed density peaks like that of the BB test \citep{BB_1979} is unlikely. However, in some cases, it is possible that fragmentation could occur and we are simply not resolving separate density peaks that are spaced closer than the size of one grid cell $\Delta x$. Thus, we aim to remove sinks that could have fragmented shortly after formation on the basis of their initial angular momentum and mass. That is, sinks are removed if all of their initial angular momentum and mass when put into circular, equal-mass binaries results in separations that would be unresolved in our simulations.
Lastly, we consider the conditions for rotational fragmentation of collapsing cores from \citet{Sterzik_2003}, which dictate that if the initial ratio of rotational energy to gravitational energy $\beta_0 \geq 0.02$ for a centrally concentrated core, the system is liable to rotationally fragment into a binary.

\section{Results}

The initially sub-virial cloud undergoes global gravitational collapse, growing sheet-like and then filamentary over time. Gaseous overdensities seeded by the decaying turbulence rapidly grow until they reach the threshold density and form sinks, preferentially embedded in the filament. 
The fiducial run  is evolved to $1.3 t_{\rm ff}$ 
(Figure \ref{fig:mainevol} shows the simulation column densities and sinks at four time snapshots), during which time it forms 115 sinks.  At 1 $t_{\mathrm{ff}}$, the median sink mass and median enclosed patch mass are 4.5 $M_{\odot}$ and 10.2 $M_{\odot}$, respectively. 

\subsection{Angular Momenta Distributions}

\begin{figure}[htb]
    \centering
    \includegraphics[width=\columnwidth]{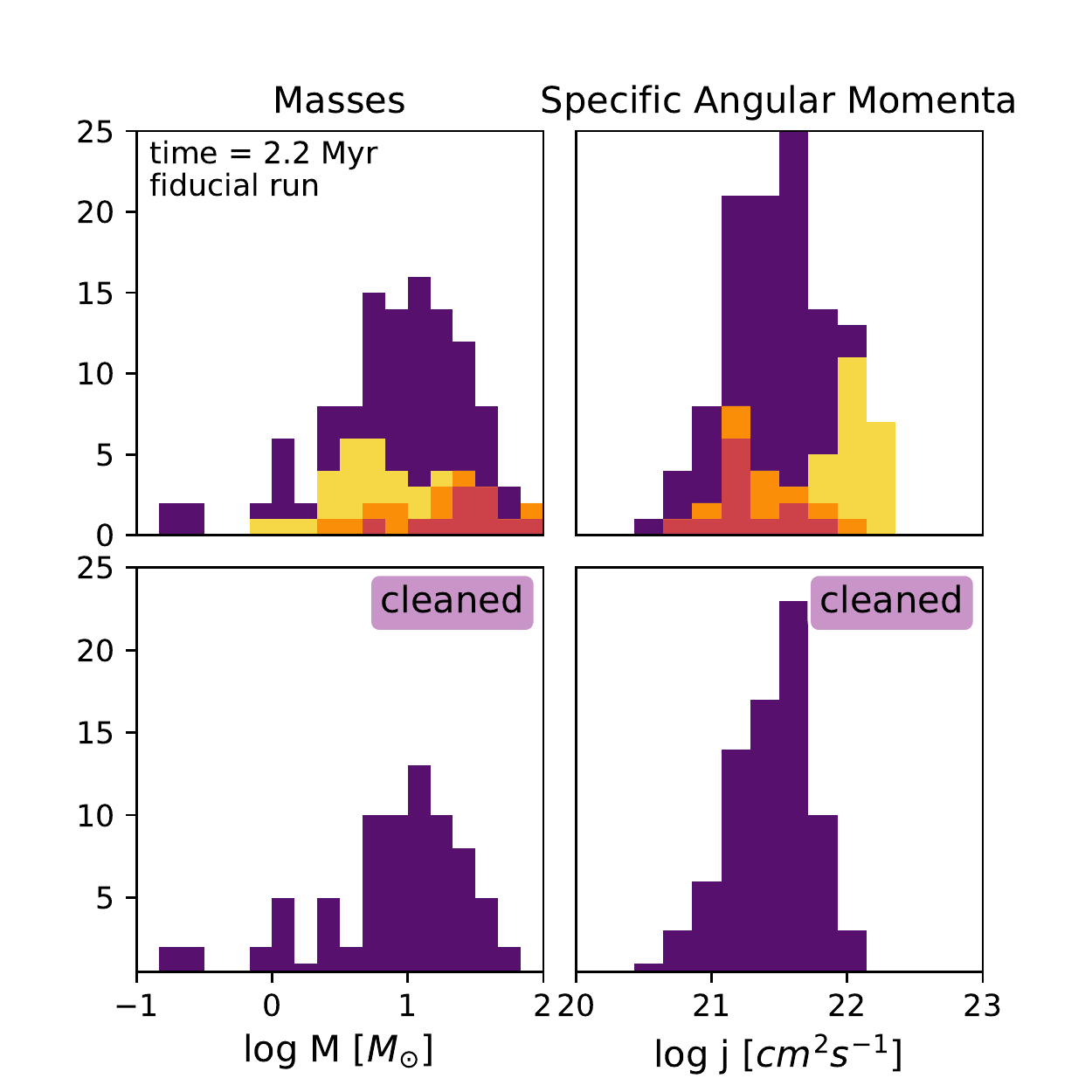}
    \caption{Top row: Stacked histograms showing the masses (left) and specific angular momenta (right) of all of the sink-patch systems in the simulation (violet), with the likely multiples highlighted: sinks that merged as "mergers" in red, sinks that would have orbital radii within one central cell as "unresolved" in orange, and sinks that could rotationally fragment as "rotational" in yellow. Bottom row: Histograms showing the distributions of sink masses (left) and patch specific angular momenta (right) within the patches in the simulation after the likely multiples have been removed. The sinks shown are for the fiducial run at a time of 2.2 Myr into the simulation. }
    \label{fig:binaries}
\end{figure}

Figure \ref{fig:binaries} shows the values of
  mass and angular momenta enclosed in the patch for the fiducial run at a time of 2.2 Myr and highlights the populations of potential contaminants from each category (e.g., likely unresolved multiple systems). We find that the specific angular momenta at all times are consistent with a log-normal distribution. At the end of the fiducial run, the median value of the specific angular momentum is  $3.2 \times 10^{21} \mathrm{cm^2} \mathrm{s^{-1}}$, or 
 $\sim 1.0 \times 10^{-2}  {\rm km s^{-1} pc}$ for the fiducial run; this is in reasonable agreement with the results of \cite{Goodman_1993} at scales of $\sim 0.05 - 0.1 $~pc. 
 
 \par
\begin{figure*}[htb]
    \centering
    \includegraphics[width=\textwidth]{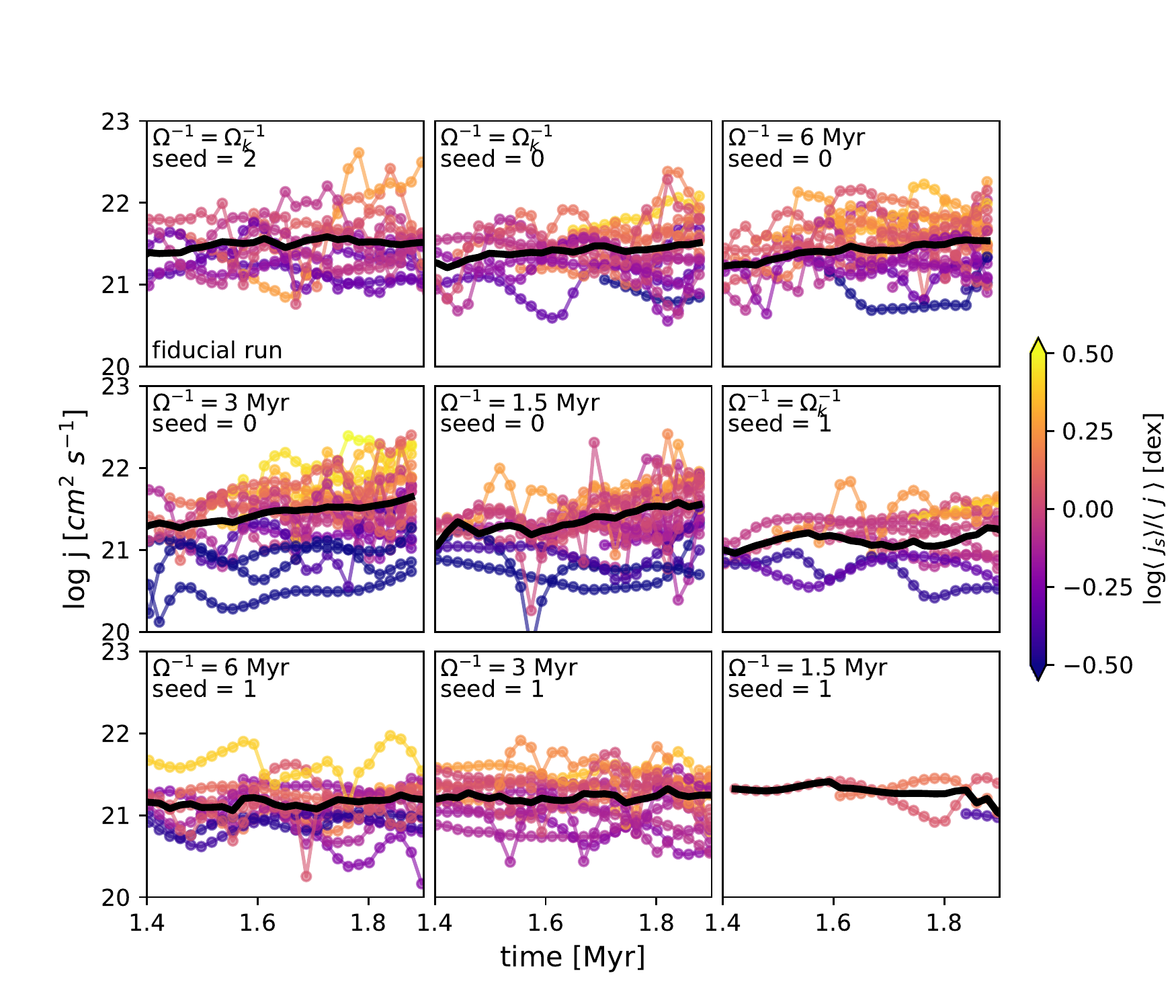}
    \caption{ Specific angular momenta of subsets of patches for each of the high resolution runs from 1.4-1.9 Myr into the simulation, covering the formation of the first sinks to 1.1 $t_{\rm ff}$ are shown. The colored tracks are individual sinks in the run and the bold black line is the median value of the specific angular momentum at each time. The colorbar denotes how far the mean value for each individual sink track deviates from the median of all sinks in dex.  These colors identify individual sinks, making it easier to see that there is little time evolution in $j$.}
    \label{fig:jtracks}
\end{figure*}

 Overall, the median specific angular momentum varies very little over time, while individual sink-patch specific angular momenta can be highly variable, as shown in Figure \ref{fig:jtracks}. These  patches can drastically increase and decrease their total angular momenta over small accretion episodes, however, the long term trend is that  patch specific angular momenta do not grow appreciably over time. During accretion episodes, patches seem just as likely to lose angular momentum as they are to gain it.
 
 \par In addition, we construct the spin parameter,  a ratio of the specific angular momentum in the sink-patch to the maximum possible angular momentum the core can have $, \lambda_s = j/(p(GMR)^{1/2})$ as a measure of the angular momentum budget in use for a patch,  where $M= M_s + M_p$ is the total mass enclosed within the patch and $R = r_p$. For specificity, we adopt $p=2/5$, the rotation coefficient for a uniform density sphere. The behavior of the median spin parameter closely follows that of the total angular momentum and is also log-normally distributed. 
 
 \par
\begin{figure*}
    \centering
    \gridline{\fig{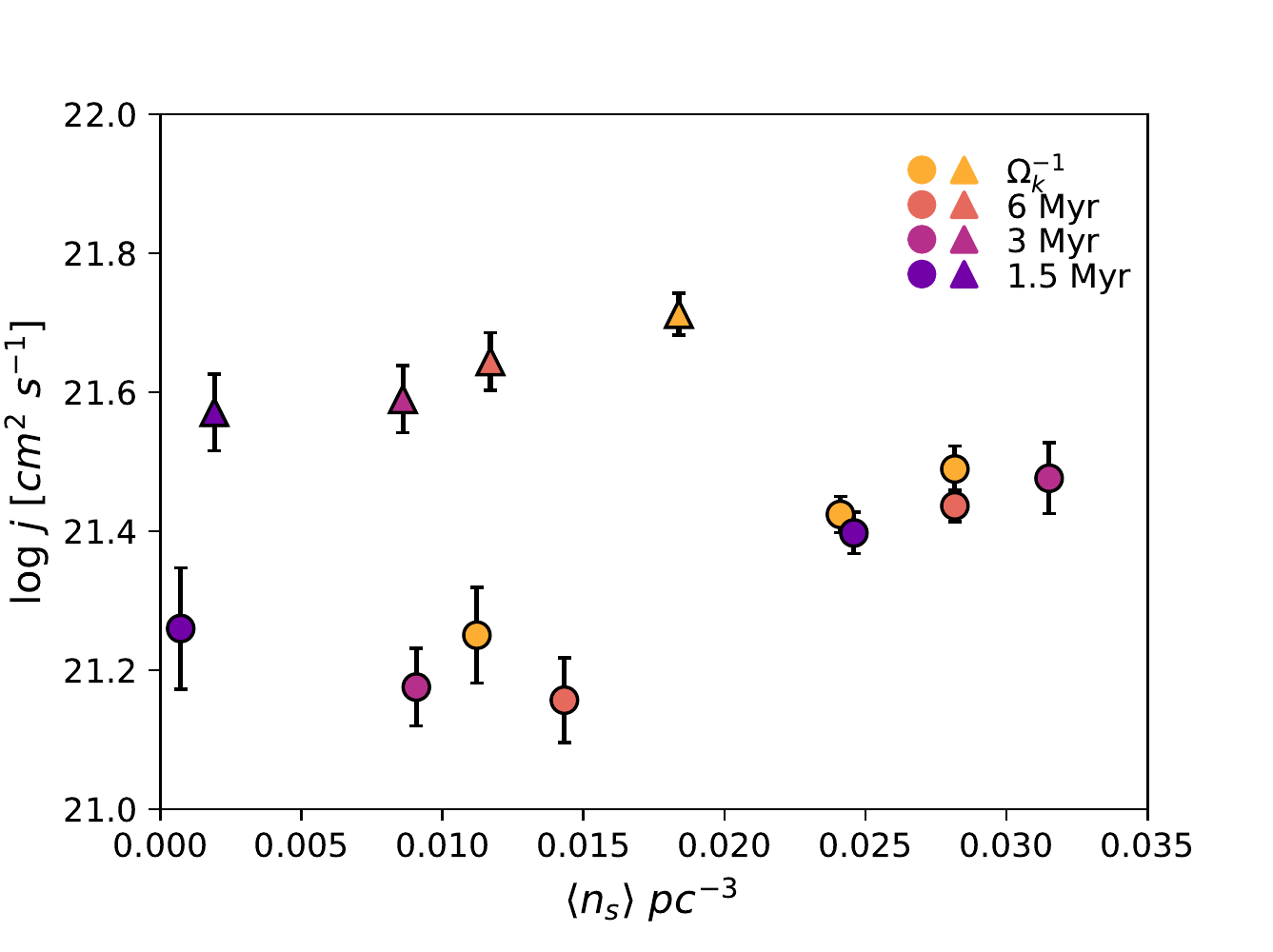}{0.5\textwidth}{(a)}
          \fig{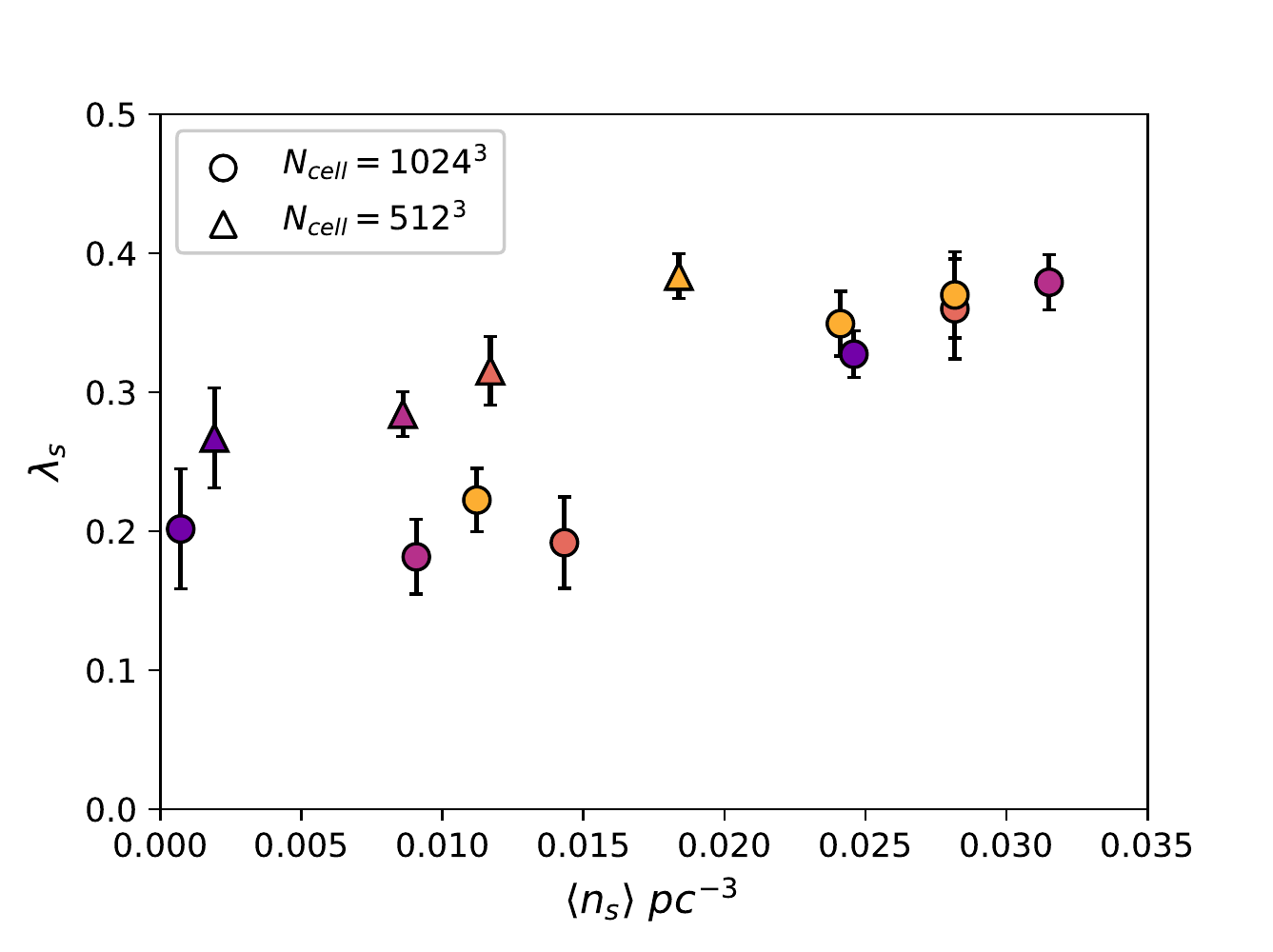}{0.5\textwidth}{(b)}}
    \caption{For 4 intermediate ($N_{cell} = 512^3$) and 9 high resolution ($N_{cell} = 1024^3$) runs, shown here are median quantities averaged over the simulation time, weighted by the number of sinks at each time for (a) specific angular momenta and (b) spin parameter  $\lambda_s$ vs the bulk stellar density (at $t_{\rm ff}$) for each simulation. Vertical error bars denote the standard deviation of the median value over time. Markers of the same color denote runs that have the same initial rotation input but different turbulent seeds; for the initial rotation inputs of $\Omega^{-1} = \Omega_k^{-1}, 6, 3, 1.5 \ \rm Myr$, the marker colors are yellow, peach, magenta, and violet, respectively. Both quantities scale with bulk stellar density, although there appears to be less correlation at lower stellar densities. The intermediate and high resolution runs are measured at different scales, $r_p = 0.1 \ \rm pc$ and $0.05 \ \rm pc$, respectively. }
    \label{fig:densityscaling}
\end{figure*}

 The median values for specific angular momentum in Figure \ref{fig:jtracks}  are not consistent between runs. In Figure \ref{fig:densityscaling}, we plot the quantities as a function of the bulk stellar density $\langle n_s \rangle$ at one free fall time for each simulation - both intermediate resolution runs where metrics are measured at $r_p  = 0.1 \rm pc$ and high resolution at $r_p = 0.05 \rm pc$. The bulk stellar density is calculated by computing the density of stars over the minimum spherical volume that contains every star in the simulation. 
 
 \par Among the high resolution runs, the median specific angular momenta at $r=0.05 \ \rm pc$ are between $\sim 1.2 - 3.2 \times 10^{21} \rm cm^2 \rm s^{-1}$ and between $\sim 4 - 5.5 \times 10^{21} \rm cm^2 \rm s^{-1} $ for intermediate resolutions measured at $r_p = 0.1 \rm pc$ (Figure \ref{fig:densityscaling} a). At similar bulk stellar densities, there is a factor of  $\sim 2 - 2.5$ offset between the specific angular momenta of the intermediate and high resolution runs. Given the two-fold increase in resolution, a factor of 2 difference in specific angular momentum is expected from basic scaling arguments. Using the spin parameter reduces this effect, shown in Figure \ref{fig:densityscaling} b where the gap between resolutions narrows. In either case, it is evident that both specific angular momenta and spin parameter directly scale with $\langle n_s \rangle$. In terms of angular momentum budget, higher stellar densities lead to $\sim 35\%$ of angular momentum budget usage, while lower stellar densities, where stars form in more isolated environments, have stars that use $\sim 20 -25\%$ of the maximum.
 \par Different bulk stellar densities can be a result of the random seed for turbulent input or the initial additional angular momentum input. The turbulent seeds contribute to different initial filament geometries; initially centrally concentrated filament geometries tend to produce a higher stellar surface density as opposed to simulations which produce a spoke of sub-filaments which form a more diffuse network of stars. 
 \par
\begin{figure*}[htb]
    \centering
    \includegraphics[width=\textwidth]{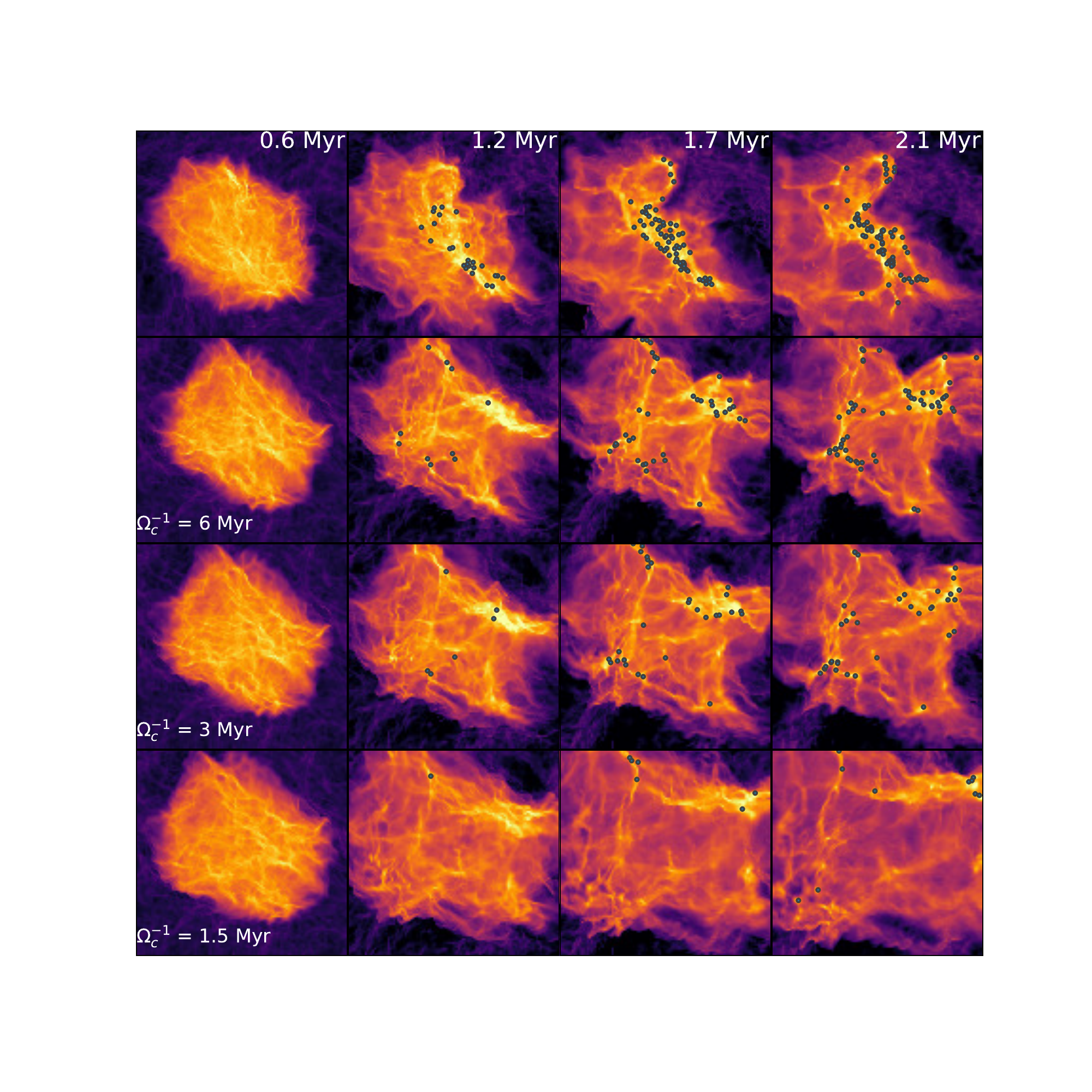}
    \caption{ Column density projections of the runs with differing initial  global rotation input zoomed in on the central 12 pc of the simulation box. Each row shows a run at $512^3$ resolution with a different amount of initial global cloud rotation, $\Omega_c$, where the first row has no initial additional bulk rotation, shown at four different times; 0.6, 1.2, 1.7, and 2.1 Myr. The column densities are normalized to the same value, with sink positions overplotted as gray dots.}
    \label{fig:columns}
\end{figure*}

 Initial angular momentum input in the form of an initial constant cloud angular velocity will spread filaments creating more diffuse areas of star formation (Figure \ref{fig:columns}). This will have less of an effect on random seeds that already produce distributed star forming filaments. For the intermediate resolution runs shown, increasing angular rotation means that fewer sink particles are formed at larger mean separations, producing less populated,  more diffuse clusters. The median masses of the sinks and patches do not vary significantly for increased angular speeds. At the highest resolutions, rotation has a less pronounced effect on the numbers of stars formed (see Table \ref{tab:allruns}), but the stars are still farther apart than for non initially rotating cases. 

\subsection{Spin Alignment}
\par
\begin{figure*}

\gridline{\fig{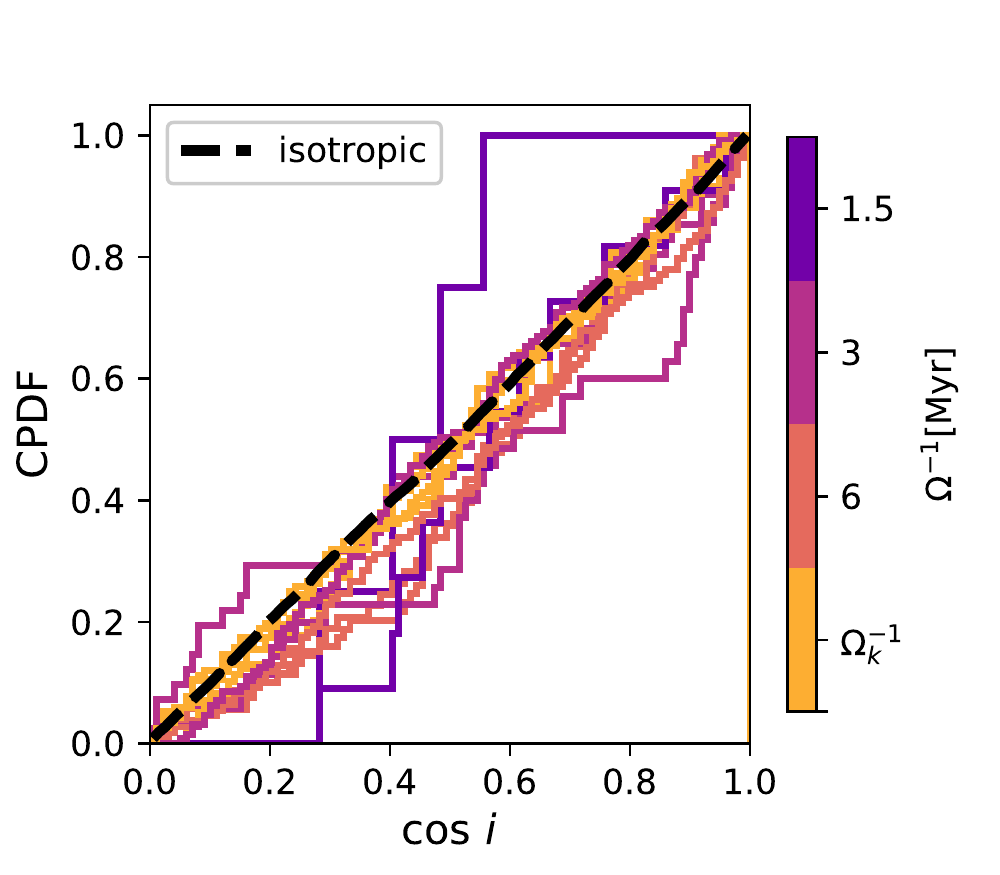}{0.5\textwidth}{(a)}
          \fig{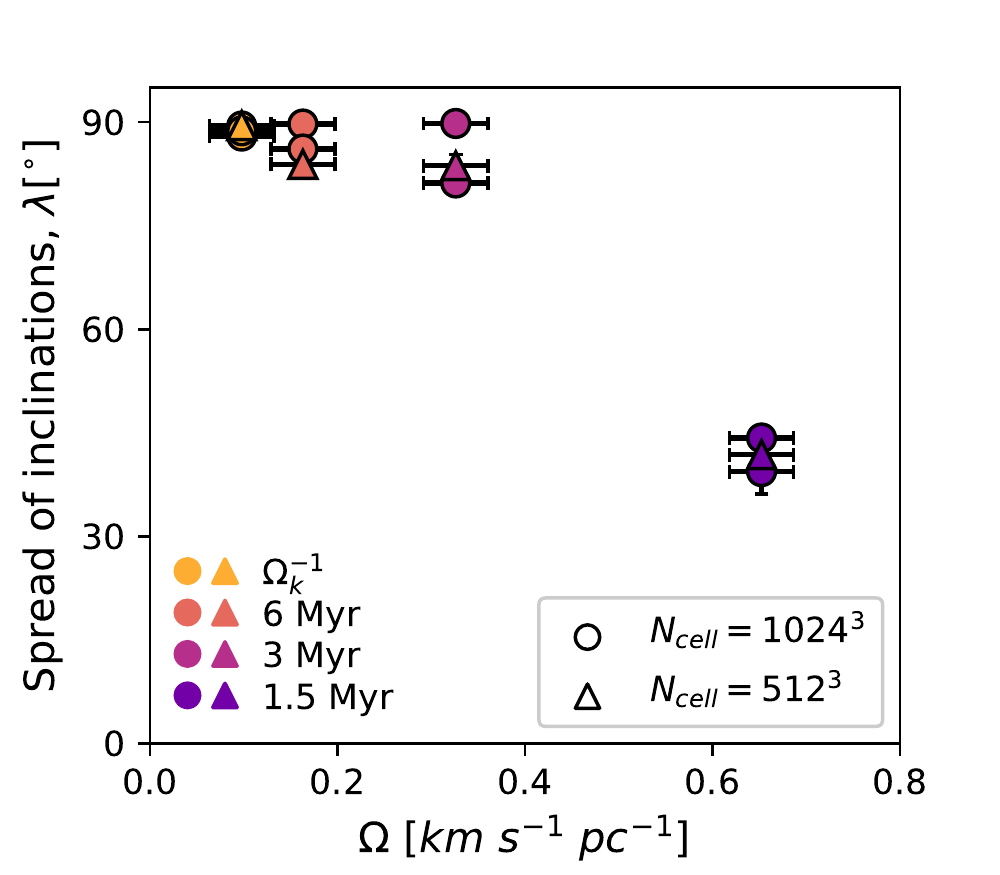}{0.5\textwidth}{(b)}}
\caption{Comparison of runs with varying amounts of initial global angular momentum.  a) The cumulative probability distribution of $\cos \ i$ for 9 selected high resolution runs compared to the expected CPDF (black dashed line) for a completely random isotropic distribution of inclinations show that most rotating clouds still appear very isotropic. The lines are color-coded according to the initial amount of angular momentum input, where the orange lines correspond to those clouds that just have the base amount of turbulent input designated $\Omega_k^{-1}$. b) The resultant fit of the spread of mutual inclinations for the different rotation inputs (color-coded to match the left panel) plotted against the initial angular speed  including contributions from the turbulence at cloud scales, $\Omega_k$}, error bars show expected level of variation from different turbulent seeds. Fitting the average CPDF across all time for the spread in mutual inclination, $\lambda$, shows that all but the fastest rotating clouds are consistent with a cone angle of $90^{\circ}$, such that most are isotropically oriented with no mutual inclination.  In summary,
significant alignment of patch inclinations only occurs for the case with the fastest global initial rotation.
\label{fig:angles}
\end{figure*}

 The initial global rotation of the cloud actually leads to modest decreases in average protostellar core angular momentum compared to its total budget on account of decreasing the stellar density of the star forming environment; the consequences for the direction of the angular momentum vector are less clear. Throughout the accretion process, the orientation of the angular momentum axes of individual patches are highly variable. As such, we look at orientations of the entire patch ensemble as a collection of inclinations to determine if the direction of initial angular momentum input could get imprinted on the cluster. In Figure \ref{fig:angles}a we show the cumulative probability distribution functions (CPDF) for $\cos i$ of 9 selected high resolution runs near the end of their simulation times, where $i$ is the inclination of the patch angular momentum axis relative to the axis of the initial angular momentum input for rotation frequencies of $\Omega_c^{-1} = $ 6, 3, 1.5 Myr and a few non-rotating runs for comparison. The non-rotating clouds have modest initial angular momentum from the injection of turbulent energy at the start, which would correspond to an eddy turnover time of $\sim 10 \,  \rm Myr$ at cloud scales. Most of the runs, even those with modest amounts of initial global angular momentum input, are fairly consistent with an isotropic distribution of inclinations, uniform in $\cos i$. {Using the K-S statistic, we can not reject the null hypothesis at a 95\% confidence level for all runs in the sample. Only cases with the largest input of global cloud rotation are marginally consistent at a 90\% confidence level, but they are also the runs that produce fewest sinks.}
\par We fit the CPDF to a model  which assumes a conical distribution of vectors about a line at an angle $\alpha$ from the assumed line of sight (which we take to be the axis of initial angular momentum input from the rotating runs) with a conical spread of $\lambda$, similar to form in \citet{Jackson_2010}. Thus, if spin axes are randomly oriented, $\alpha = 0, \lambda = 90^{\circ}$. Using the python package emcee \citep{emcee_2013}, we fit the average CPDF over the entire run time of simulations to models of $\cos i$. The models are generated according to the methods in \citet{Jackson_2010}, where the final model is of the form $\cos i = \sin \alpha \sin(\cos^{-1}(1 - R (1 - \cos \lambda)) \cos \phi + \cos \alpha (1 - R(1-cos\lambda))$ where $R$ and $\phi$ are drawn from random distributions within $[0,1]$ and $[0,2\pi)$, respectively.
\par In Figure \ref{fig:angles}b, we show the results of the fits for the $\lambda$ parameter for both the high and intermediate resolutions runs from which it is evident that for all but the most rapidly rotating clouds, the spread in distribution, $\lambda$ is consistent with $90^{\circ}$ and favors no mutual alignment among the cores. The fastest rotating cloud's speed is near break-up, where $1.5 \, \rm Myr^{-1} \sim 0.65
\, \rm km \, s^{-1} \, pc^{-1}$, which is considerably larger than typical values 
$\sim 0.05 - 0.2
\, \rm km \, s^{-1} \, pc^{-1}$ for Milky Way giant molecular clouds \citep{imara11}. While the $\cos i$ distributions remain nearly uniform for most clouds, the amount of retrograde orbits decreases with added rotation, where retrograde in this case refers to a core's rotation axis being opposite that of the initial rotation axis of the global cloud. This is consistent with a slight favoring of alignment with the global rotation axis, but would not be obvious from observations of just the inclinations.

\section{Discussion}
\par In the following we use our patches as proxies for protostellar cloud cores.
While our patches are defined by fixed grid sizes rather than some density criterion, these sizes are comparable to those of cores in low-mass star-forming regions, and we showed in \cite{Kuznetsova_2018a} that patch masses are reasonable approximations for the self-gravitating mass surrounding sinks.

\par  The median values of the specific angular momenta are consistent with those derived from measurements of velocity gradients in protostellar core observations at similar size scales, where at  radii 0.05-0.1 pc the specific angular momentum from \cite{Goodman_1993} would be on the order of 0.002-0.015 km $s^{-1}$ pc or $0.6-4.6 \times 10^{21}$ $\mathrm{cm^2}$ $\mathrm{s^{-1}}$, compared to $j \sim 3.2 - 5 \times 10^{21} \mathrm{cm^2} \mathrm{s^{-1}}$ at the patch edge in our fiducial and intermediate resolution run.  These comparisons yield a good agreement, with varying uncertainties in the observations likely due to projection effects \citep{zhang18}. 

\par With reference to the distribution of angular momentum directions, there are numerous ways to infer the spin axis orientation of stars e. g. the orientation of outflows/jets \citep{Stephens_2017} or measuring both the period $P$ and line of sight $v\sin i$ of magnetically active late-type stars \citep{Jackson_2010}. \citet{Jackson_2010} posit that observations of mutual alignment of inclinations could yield information about the initial rotation of the star forming region. Yet, our results in Figure \ref{fig:angles} show that unless the initial rotation is incredibly high, we would not expect to see obvious trends in the alignment of specific angular momentum axes. These results are fairly consistent with findings from \citet{corsaro_2017} and with the mostly random distribution of inclinations found in many star cluster and star forming regions \citep{Jackson_2010, Menard_2004}. 

\par Using the spin parameter, we find that core angular momenta in our simulations are on average  $25-40\%$ of the maximum possible, and thus are dynamically important. With the core masses, radii, and specific angular momenta from Figures 1 and 5 in \citet{Offner_2008} and performing the same procedure to compute the spin parameter for their cores, we find that the median value of the spin parameter at all scales for the undriven (as is the case in our study) turbulent case is $0.28$, consistent with the low end of our simulations. The driven turbulent case creates cores that have far less of their angular momentum budget at small scales, with a median spin parameter of $0.07$ across all scales. At our intermediate patch scale ($\sim 0.1$ pc), the values of the spin parameter and angular momenta of both the undriven and driven cases in \citet{Offner_2008} are both consistent with our distributions. However, at higher resolution for patch sizes of $0.05 \ \rm pc$, the driven turbulence case has a median $\lambda_s \sim 0.1$, diverging from our results, the effects of driven turbulence start becoming significant at core scales $r_p \lesssim 10^4 \ \rm AU$.

\subsection{Local Generation of Angular Momentum}

\par We suggest that the best explanation of where cores get their angular momentum that explains the results of the parameter study is one in which gravitational interactions between gas overdensities, including dense cores designated by the sink+patch construction,  impart the initial angular momentum to the core.  
Our argument in favor of local generation of angular momentum rests on two observations that make inheritance of
initial angular momentum unlikely: one, that the directions of core angular momenta are not correlated with any specific direction, including that of the initial global angular momentum of the cloud (Figure \ref{fig:angles}); and two, that the specific angular momenta increase with increasing density of sinks ( and thus closer separations) (Figure \ref{fig:densityscaling}) , as one might qualitatively expect for local torquing. There is more than enough force in gravitational interactions for small scale overdensities to create torques; and if the overall spatial density of sinks decreases and density fluctuations were smeared out, one could expect lower values of angular momentum as gravitational interactions must act over a greater distance.

\par The results of varying the initial cloud rotation (Figure \ref{fig:columns}), show that effect of varying the initial global cloud angular velocity $\Omega_c$  is the total number and spatial density of the gas and sinks, where the global cloud rotation tends to shear out a portion of the gaseous overdensities, producing more diffuse, less embedded clusters. The angular momentum content of the sinks depends on the star forming environment rather than the initial angular momentum content of the cloud.  Increasing the global angular momentum content of the cloud does not pass down to the sinks; in fact,  the average angular momentum content decreases with increased initial cloud rotation. In Figure \ref{fig:densityscaling} for example, the intermediate resolution runs with $N_{cell} = 512^3$ are arranged in the order of their initial cloud angular momentum, with the lowest $\langle n_s \rangle$ corresponding to the highest initial rotation speed.

\par As a caveat, similar amounts of rotation input will affect the stellar densities of clusters differently depending on the resolution and geometries generated by the random seed for the turbulence.  Higher resolution runs produce more sinks as the amount of resolved overdensities will increase in an isothermal simulation such as ours. However, increased initial global rotation will still produce more diffuse clusters and filaments as it slows the collapse of the cloud. The local filamentary environment seems to play an important role in the properties of protostar populations, particularly when it comes to the starting ingredients of protostellar disks. 

 \par Our construction of the spin parameter is based on the metric used for characterizing the angular momentum content of cold dark matter haloes in cosmological simulations, where the spin parameter is the ratio of the angular momentum to the maximum amount possible at the virial radius of the halo.  On the side in favor of local angular momentum generation, we make an analogy with dark matter simulations \citep[][e.g.]{Bullock_2001}
which argue that a combination of tidal torques and mergers produce the halo spin parameter distributions, which are log normal just as ours are.

\par  The (initial) injection of turbulence at small scales could be thought to be responsible for imparting the initial angular momentum to the cores. However, the nature of the turbulent power law power spectrum, such as the $P(k) \propto k^{-4} dk$ used here, means that there will inherently be less energy and angular momentum at smaller scales. A conservative order of magnitude calculation of the angular momentum of an eddy on the patch scale $\sim 0.05 - 0.1$ $ \mathrm{pc}$, assuming that the eddy frequency $\Omega_k \propto k ^{-1/2}$, where the largest scale fluctuation frequency matches that of an eddy with a speed of Mach 8, $\Omega_k^{-1} \sim 10$ $ \mathrm{Myr}$, results in an angular momentum $j = 2/5 R^2 \Omega \sim 10^{20} - 10^{21} $ $\mathrm{cm^2} \mathrm{s^{-1}}$, which is roughly an order of magnitude below the minimum angular momentum seen in the fiducial run, thereby not enough to generate the typical values of the core angular momentum in the simulation. In addition, runs across the parameter space all have the same turbulent energy initially injected, thus, the angular momentum across the runs of varying global cloud rotation speed should, at the very least, remain fairly constant if  angular momentum were directly inherited from the turbulent eddies. 

\par In contrast, \cite{burkert_2000} and \cite{chen_2018} argue that the core angular momenta are inherited from turbulence.  In the former case, the absence of self-gravity (A. Burkert, personal communication) means that the imposed turbulent velocity field is the only possible source of angular momentum.
In the latter case, the small scale (1 pc) of the simulation requires the imposition of prescribed inflows and turbulent motions \citep[see][]{chen_2015}
which in our case arise naturally from gravitational collapse from larger scales, seeded by overdensities created by the rapidly-decaying initial turbulence.  Our simulations may also exhibit stronger gravitational driving because \cite{chen_2015,chen_2018} evaluate core properties at the onset of collapse of the most 
evolved core, whereas we can follow the evolution through sink formation and accretion.

\subsection{Implications for Disk Formation}
\begin{figure*}[h!]
\centering
\begin{minipage}{.32\paperwidth}
  \centering
  \includegraphics[width=\columnwidth]{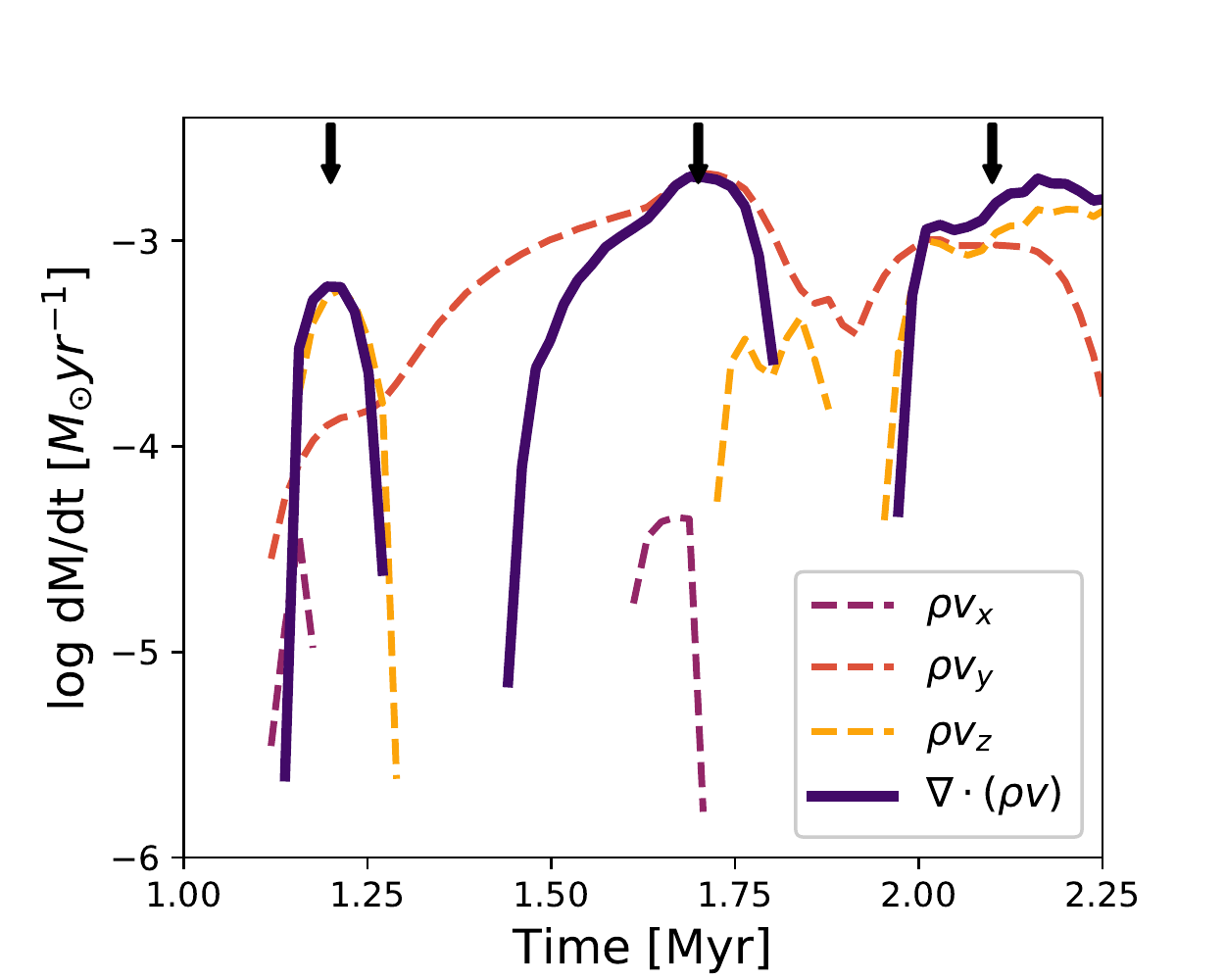}
  \caption{Net mass flux into the patch in three orthogonal directions (dashed lines) and the total mass flux into the patch (violet line). The magnitude of the mass flux plotted is normalized such that the integrated amount under the curve matches the patch mass at the end of the simulation in order to account for material that does not become bound to the sink and leaves the patch. The arrows annotate specific times of high inward mass flux at 1.2, 1.7, and 2.1 Myr that are shown as column snapshots in Figure \ref{fig:snapshots}. Note that material at each high flux event primarily comes in from different directions.}
  \label{fig:mflux}
\end{minipage}%
\begin{minipage}{.48\paperwidth}
  \centering
  \includegraphics[width=\columnwidth]{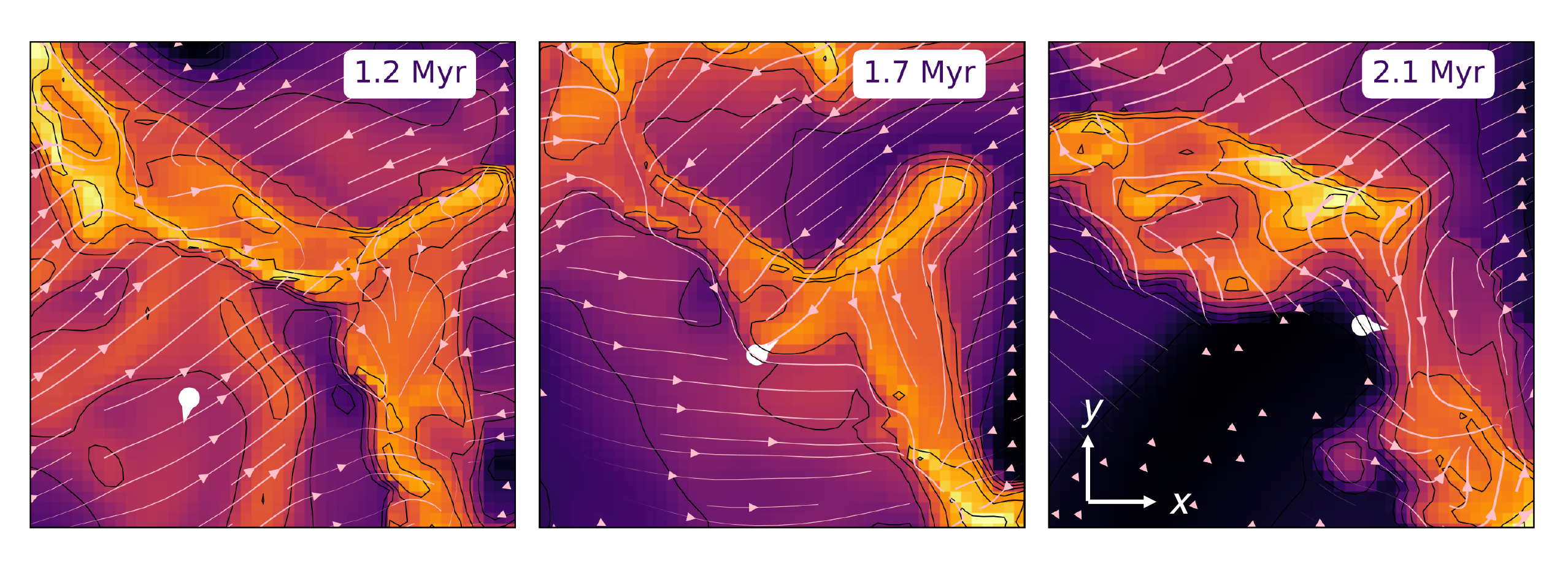}
  \caption{Column densities summed over 0.5 pc of the same 2 pc $\times$ 2 pc region taken at the times corresponding to the high mass flux events of the sink-patch system annotated in Figure \ref{fig:mflux}; 1.2, 1.7, and 2.1 Myr. The sink-patch system is shown as a white marker with an arrow indicating its motion in the plane of the image, the background streamlines show the flow of material within the region. During the three frames, the sink's accretion events are at first dominated by flow into the filament, then flow along the filament as the sink becomes embedded within, and finally the flow directed toward the sink as the sink becomes large enough to deplete some of the filament gas. The surrounding environment is heterogeneous -- lumpy, and accretion is dominated by these irregularities.}
  \label{fig:snapshots}
\end{minipage}
\end{figure*}

\begin{figure*}[b]
    \centering
    \includegraphics{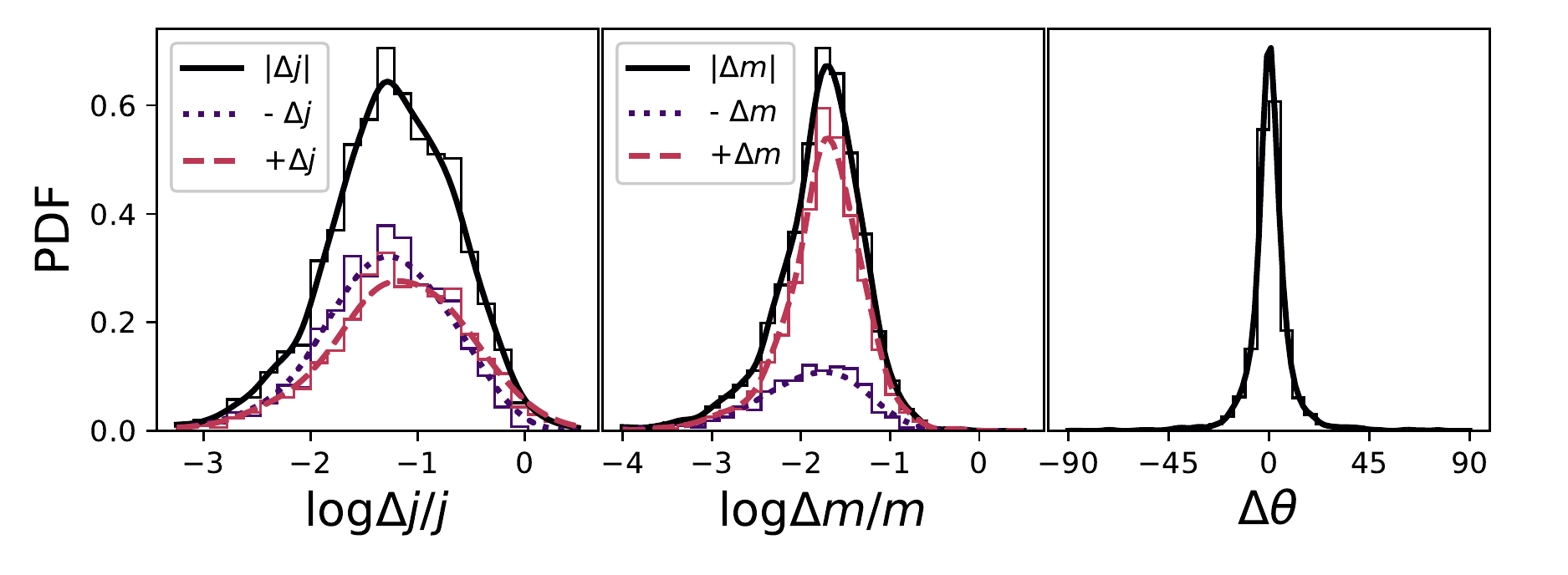}
    \caption{From left to right: Probability density functions of relative changes in patch specific angular momentum, mass, and direction of angular momentum between data outputs for the intermediate resolution run shown in Figures \ref{fig:mflux} and \ref{fig:snapshots}, at a cadence of 0.02 Myr between outputs. Left: The magnitude of the relative change in angular momentum (black solid line) is a skewed Gaussian centered at $\lesssim 10 $ \% of the previous angular momentum. The pink dashed line represents increases in the angular momentum while the violet dashed line represents total decreases in the angular momentum. Center: The magnitude of the change in mass peaks at $ \sim 0.02 $ \%, but most changes are accretion events rather than material flowing out of the patch. Right: Change in total angular momentum direction is centered at 0, as changes in direction require large relative changes in specific angular momentum. }
    \label{fig:model}
\end{figure*}

\par Angular momentum is a requirement for disk formation and many models of disk formation and evolution will assume an accretion of angular momentum by the protostar from the surrounding cloud like in \citet{TSC_84}, where rotating collapse grows a disk by depositing material at the centrifugal radius, $r_c = j^2 / GM$.  For simplicity, these models assume a constant angular velocity cloud from which angular momentum is inherited as the accretion radius of the protostar grows and material falls in from farther away. The angular momentum then depends directly upon the angular velocity of the cloud. Using the TSC model as an example framework, these assumptions will lead the specific angular momentum to grow like $j\propto t^2$, or the spin parameter $\lambda_s \propto t^{3/2}$. 
 However, seen in Figure \ref{fig:jtracks}, the mean of the sink specific angular momenta does not significantly change with time.
 The lack of change in sink angular momentum content suggests that mass infall to disks will weight outer regions more heavily than in the TSC model. Previous studies show that infall to larger disk radii may also help trigger gravitational instabilities \citep[e.g.,][]{zhu12}.

\par The accretion of angular momentum in our simulations is not a smooth monotonic process. This already challenges the standard ideas of what the initial conditions for disk formation and evolution should look like. We conduct a case study by looking in detail at the dynamics of the gas entering and leaving one particular sink-patch system. In Figure \ref{fig:mflux}, we show the mass flux entering the patch through the individual faces of the patch and compare it to the total inward mass flux across the patch. At times, there is no net mass flux into the patch at all and the accretion has an episodic quality, where several different epochs of mass flux inward can be identified in Figure \ref{fig:mflux}, annotated with arrows marking the timestamps of the snapshots plotted in Figure \ref{fig:snapshots}. 
\par In this example, it is possible to connect the interactions of the sink-patch system with its environment to specific accretion epochs.  In the first epoch, at 1.2 Myr into the simulation, the sink accretes from a flow of gas falling into a filament, however, by 1.7 Myr, the sink and filament have come together and the sink is embedded while gas flows along the filament, onto the sink. The filament itself is not a smooth object, but quite lumpy. At 2.1 Myr, the sink has quickly accreted a large clump from the filament and is accreting the next nearby overdensity. This example demonstrates how a dynamic clump-filled environment in which gravity dominates the dynamics naturally lends itself to episodic core accretion, in which infall onto the core is not isotropic, but dominated by the direction of flows in the environment. 
\par In our case, non-isotropic infall can explain why the specific angular momentum of sinks does not evolve over time. In Figure \ref{fig:model}, we plot the relative changes in the magnitude and direction of the specific angular momentum, as well as the relative changes in the total mass enclosed in the patch for sinks between outputs taken every 0.02 Myr as probability density functions. Looking at the changes in angular momentum for the cores, it is easy to see why the total angular momentum of the cores does not show smooth growth over time -- cores are about just as likely to gain angular momentum as they are to lose it. This type of behavior is a natural consequence of episodic filamentary, non-isotropic accretion we show in Figure \ref{fig:mflux}, most commonly when material is accreted from many different directions such that the mass of the patch grows, the direction of the angular momentum vector for the sink changes, but the total sink angular momentum will fluctuate about a value.

\par The simple picture in which smooth spherical infall of rotating material builds up a disk may require revision. A more time-dependent, filamentary-type of infall may have implications for making disks more susceptible to gravitational instability from episodes of mass-loading.

\section{Summary}
\par

Our simulations of protostellar core
formation in a globally-collapsing
molecular cloud show that the
angular momenta of individual cores
are not strongly affected by global cloud rotation.  Instead, the angular
momenta appear to be generated by local torques from other cores and density concentrations.
Unlike often-used models for protostellar collapse, the external
medium that can be accreted is
generally not in solid body rotation
around the central sink, resulting
in nearly constant specific angular momenta as the protostellar core
(sink-patch in our calculation) grows.
As seen in other simulations such as those of \cite{Bate_2018}, we find considerable time variability in the accretion of the angular momenta and mass, which is the product of a lumpy episodic non-isotropic accretion of material onto the protostellar cores.

The results presented in this
paper constitute the first stage of our program, which next will consider the effects of magnetic fields, with the ultimate goal of
developing more physically-realistic angular momentum distributions that can be used as inputs to models of protostellar and protoplanetary disk formation and evolution.

\acknowledgements
This work was supported by NASA grant NNX16AB46G and by the University of Michigan, and used computational resources and services provided by Advanced Research Computing at the University of Michigan, Ann Arbor and the University of North Carolina at Chapel Hill Research Computing group.

\bibliographystyle{aasjournal}
\bibliography{biblio}

\end{document}